# Conjugate metamaterials and the perfect lens

Yadong Xu, Yangyang Fu, Lin Xu, and Huanyang Chen*

*College of Physics, Optoelectronics and Energy & Collaborative Innovation Center of Suzhou Nano Science and Technology, Soochow University, No.1 Shizi Street, Suzhou 215006, China*
*chy@suda.edu.cn

**In this letter, we show how transformation optics makes it possible to design what we call conjugate metamaterials. We show that these materials can also serve as substrates for making a subwavelength-resolution lens. The so-called "perfect lens", which is a lens that could focus all components of light (including propagating and evanescent waves), can be regarded as a limiting case, in which the respective conjugate metamaterials approach the characteristics of left-handed metamaterials, which have a negative refractive index.**

It has been shown that left-handed metamaterials (LHMs)[1], which have both their permeability and permeability equal to -1, may be used to make a perfect lens: they can amplify evanescent waves, which play an important role in the resolution of the image[2]. The conventional lens has a diffraction limit: the maximum resolution of the image is always smaller than the working wavelength. The reason is the evanescent waves, that contribute to higher resolutions of the image, will exponentially decay during the propagations. A perfect lens can thereby defeat the diffraction limit and even achieve an infinitely large resolution of the image by collecting all components of waves, both propagating and evanescent. Many proof-of-principle experiments have demonstrated negative refraction in LHMs[3, 4, 5, 6, 7, 8]. However, to implement such metamaterials experimentally, resonant components should be utilized, which will naturally incur some losses, compromising the imaging capability. It has been suggested that adding gain to the metamaterials could improve their functionality[9, 10, 11]. Recently, parity-time (PT) symmetric metamaterials have attracted a lot attention and demonstrated novel functionalities, such as unidirectional invisibility phenomena[12], coherent perfect absorption[13, 14], nonreciprocity of light propagation[15] and so on. These metamaterials, by spatially modulating loss and gain, possess a complex refractive index profile with $n(r) = n(-r)^*$, therefore demonstrating a real eigenvalue spectra. Hence, metamaterials that exhibit gain warrant more research from both theoretical aspects and experimental perspectives.

Here we will introduce the concept of a kind of metamaterial that has not been systematically studied before[16]. We would like to call them conjugate metamaterials (CMs) as the products of their permeability and permeability are positive real numbers and refractive indexes are well defined. It has been shown that non-attenuated propagation of electromagnetic (EM) waves is possible in these CMs in Ref. [16]. However, neither clear physics picture nor novel functionalities of these metamaterials has been proposed in literature. Just like LHMs with both their permeability and permeability equal to -1 has been proposed by Veselago in 1960s, yet its property of perfect lens was revealed in 2000 by Pendry. Here in this letter, we will show that transformation optics[17, 18, 19] can help us to derive this kind of metamaterial with an intuitive physics picture. At the same time, we surprisingly find some novel properties, especially in terms of LHMs and perfect lenses[1, 2].

Let us start from Maxwell's equations at a fixed frequency $\omega$:

$$\nabla \times \vec{E} = -i\omega\mu_0\mu\vec{H} \text{ and } \nabla \times \vec{H} = i\omega\varepsilon_0\varepsilon\vec{E}, \qquad (1)$$

and perform the following phase transformation to the electric field and magnetic field (Fig. 1a):

$$\vec{E}' = e^{i\alpha}\vec{E} \text{ and } \vec{H}' = e^{i\beta}\vec{H}. \qquad (2)$$

To obtained the above solutions, we need to modify the permeability and permeability[18, 19], i.e.,

$$\mu' = e^{i(\alpha-\beta)}\mu \text{ and } \varepsilon' = e^{-i(\alpha-\beta)}\varepsilon. \qquad (3)$$

If $\mu$ and $\varepsilon$ are originally positive real numbers, we can define a real refractive index $n$ according to

$$n^2 = \varepsilon\mu = \varepsilon'\mu'. \qquad (4)$$

These metamaterials may contain lossy media and gain media simultaneously, yet with a well defined refractive index. Here in the letter we will reveal a very important property of CMs, in particular relating to a perfect lens.

Suppose $\varepsilon = \mu = 1$, and without loss of generality, $\beta = 0$; then we have

$$\mu' = e^{i\alpha} \text{ and } \varepsilon' = e^{-i\alpha}. \qquad (5)$$

Such CMs are in fact more general than LHMs, to which they reduce for the special case of $\alpha = \pi$, in which case $\mu' = \varepsilon' = -1$ (Fig. 1b). In principle, these CMs will bring in an arbitrary phase shift for EM fields. This physics picture is straightforwardly obtained from above transformation optics.

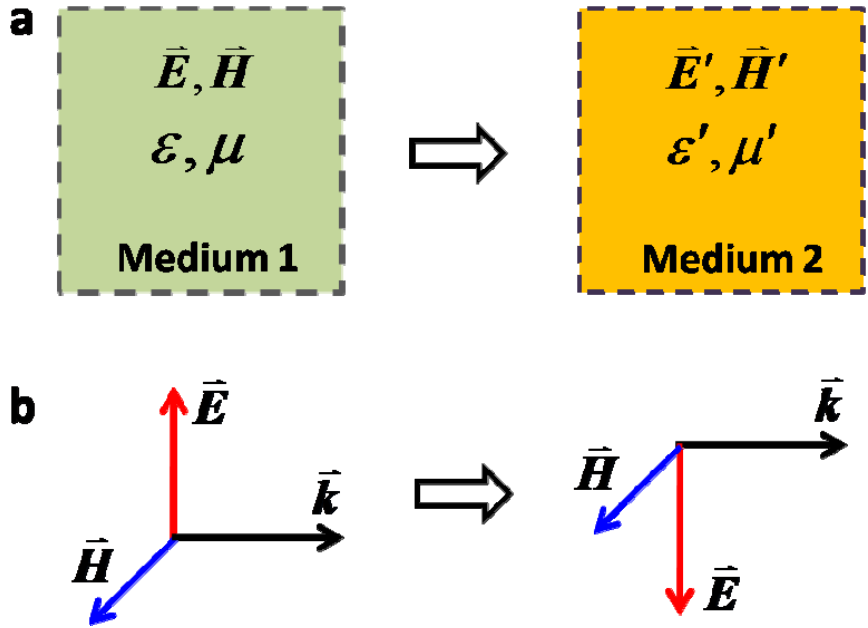

Fig. 1 (a) Phase transformation from one space to another. (b) When $\alpha = \pi$, $\vec{E}' = -\vec{E}$ (Eq. (2)). Therefore, such a transformation changes the handedness as the electric field, magnetic field, and propagating wave vector form a left-handedness.

For simplicity, we will confine $\alpha$ from $0$ to $\pi$, which means that $\mu'$ contains lossy elements while $\varepsilon'$ contains gain elements. If one exchanges their roles, the conclusions remain the same. In Fig. 2a, $\mu'$ is located in the upper part of the unit circle (red dashed half circle) while $\varepsilon'$ is in the lower part (blue dashed half circle). When $0 \leq \alpha < \frac{\pi}{2}$, the EM wave incident from air will not change its direction but will continue to propagate in CMs to the right (Fig. 2c). When $\frac{\pi}{2} < \alpha \leq \pi$, the EM wave incident from air will undergo negative refraction (Fig. 2b), as evident from a "handedness analysis" like that in Fig. 1b. For $\alpha = \frac{\pi}{2}$ (corresponding to $\mu' = i$ and $\varepsilon' = -i$), both directions are possible, so the outcome is ambiguous. Below we will see that this value corresponds to a transition point. For all the above cases, reflection should also be considered, because of impedance mismatching at the CM-air interface, except for two specific situations: when the CMs are equivalent to air ($\alpha = 0$) or when they become LHMs ($\alpha = \pi$). For more detailed calculations and explanation, see the supplementary material.

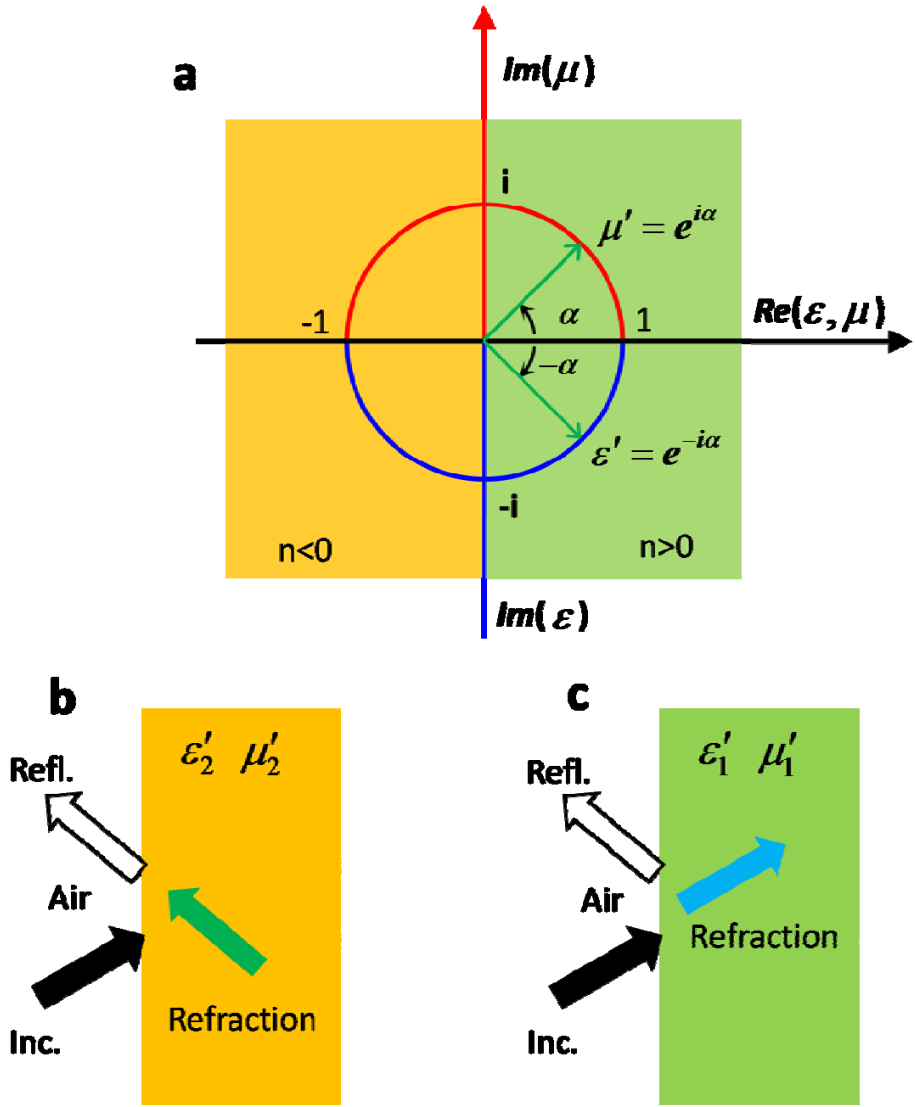

Fig. 2 (a) For $0 \leq \alpha \leq \pi$, a lossy $\mu'$ and an $\varepsilon'$ exhibiting gain correspond to the unit circle. (b) For $\frac{\pi}{2} < \alpha \leq \pi$, negative refraction happens at the interface of air and the CMs. (c) For $0 \leq \alpha < \frac{\pi}{2}$, the propagation direction of the EM wave will not change as it passes from air to the CM.

These metamaterials can also play roles in designing a perfect lens, or at least a lens with subwavelength resolution (also called as a "superlens" by Pendry [2]). For example, consider a CM slab in air. Fig. 3a shows the refraction field pattern for an incident TE plane wave for CMs with $\alpha = \frac{\pi}{4}$. The propagation direction of the EM wave is unchanged except for a small amount of reflection due to impedance mismatching at the interfaces. For $\alpha = \frac{3\pi}{4}$, there is negative refraction at both interfaces of the CM slab (Fig. 3c). For $\alpha = \frac{\pi}{2}$, the positive and negative refraction have the same ratio in the CM slab, which can be seen from the interference pattern (Fig. 3b).(See supplementary material for more details). We only show the related amplitude of the positive refracted wave and that of negative refracted wave for TE polarizations. For the positive component, the amplitude is

$$A_p = \frac{\cos\frac{\alpha}{2}}{|\cos^2\frac{\alpha}{2} + e^{2i\phi}\sin^2\frac{\alpha}{2}|}, \qquad (6)$$

while for the negative one it is

$$A_n = \frac{\sin\frac{\alpha}{2}}{|\cos^2\frac{\alpha}{2} + e^{2i\phi}\sin^2\frac{\alpha}{2}|}, \qquad (7)$$

where $\phi = k_z d = \frac{\omega}{c} d \cos\theta$ is the phase change perpendicular to the slab, $d$ is the thickness of the CM slab, $\theta$ is the incident angle from the air, $c$ is the velocity of light in vacuum, and $k_z$ is the wavevector component perpendicular to the slab.

When $\alpha = 0$, the slab becomes air, and therefore $A_p = 1$ and $A_n = 0$. When $0 < \alpha < \frac{\pi}{2}$, $A_p > A_n$, the negative component is weaker and comes from the reflection. When $\alpha = \frac{\pi}{2}$, $A_p = A_n$, so that both the positive and negative components have the same ratio, consistent with the field pattern in Fig. 3b. When $\frac{\pi}{2} < \alpha < \pi$, $A_p < A_n$, the negative component plays a more important role, while the positive component can be regarded as the contribution from reflection. When $\alpha = \pi$, the slab becomes a perfect lens, so that $A_n = 1$ and $A_p = 0$.

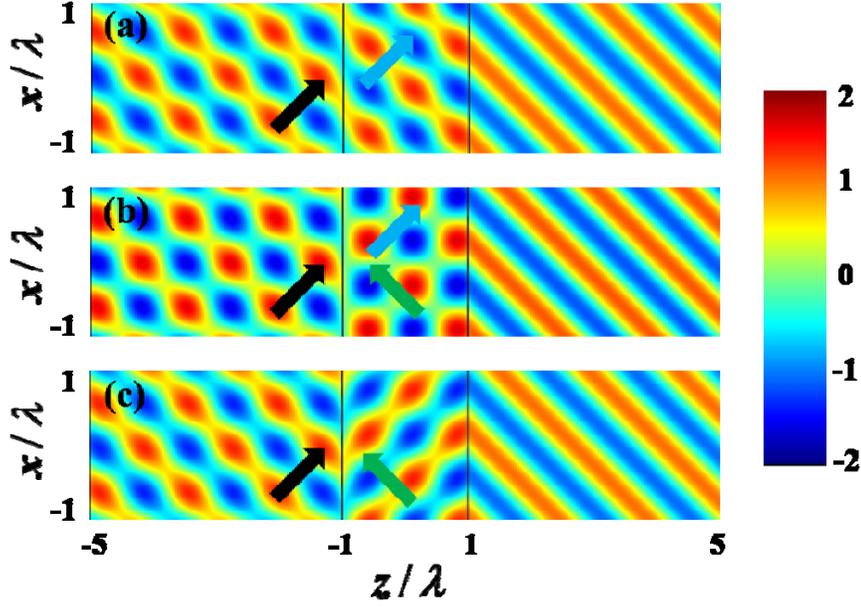

Fig. 3 The field patterns for an obliquely incident TE plane wave for CMs with (a) $\alpha=\frac{\pi}{4}$; (b) $\alpha=\frac{\pi}{2}$; and (c) $\alpha=\frac{3\pi}{4}$. The thickness of the slab is $2\lambda$ ($\lambda$ is the working wavelength). The incident angle is $\frac{\pi}{4}$.

Now we come to see whether such a CM slab can also amplify the evanescent waves and serve as a sub-wavelength lens or even a perfect lens. Following Eqs. (21) and (23) in Ref. [2] (see also in Eqs. (S17) and (S21) in supplementary materials), the amplification for evanescent waves for both TE and TM polarizations can be written as

$$T_{s(p)} = \frac{\exp(ik_z d)}{\cos^2\frac{\alpha}{2} + \sin^2\frac{\alpha}{2}\exp(2ik_z d)}. \quad (8)$$

For evanescent waves, $k_z$ can be written as an imaginary number $i\kappa$, and we can define $\Delta = \kappa d$. Then we have,

$$T_{s(p)} = \frac{\exp\Delta}{\exp 2\Delta \times \cos^2\frac{\alpha}{2} + \sin^2\frac{\alpha}{2}}. \quad (9)$$

We plot $T_{s(p)}$ as a function of $\Delta$ for different $\alpha$ in Fig. 4. When $\alpha < \frac{\pi}{2}$, $T < 1$ for all the

$\Delta$ (see the curves with triangular data points in Fig. 4a), which means that CMs with $\alpha < \frac{\pi}{2}$ cannot amplify the evanescent waves. However, for $\alpha > \frac{\pi}{2}$ there will be peaks in each curve. For example, when $\alpha = \frac{3\pi}{4}$ the peak is at around ($\Delta = 1$, $T = 1.5$), see the curve with square data points in Fig. 4a. Evanescent waves with $0 < \Delta < 2$ can then be amplified. So the CM can now serve as a subwavelength-resolution lens. When $\alpha$ goes to $0.9\pi$, more evanescent waves can be amplified and with stronger amplitudes (the peak now exceeds 3, and the range of $\Delta > 1$ is from [0, 4], see the curve with circular data points in Fig. 4a). Therefore, as $\alpha$ becomes closer to $\pi$, the resolution of the CM slab will improve.

However, no matter how close $\alpha$ is to $\pi$, there will still be a peak, and the amplification will eventually drop to below one and then to zero. We show the results for $\alpha = 0.99\pi$, $0.999\pi$, $0.9999\pi$, and $0.99995\pi$ in Fig. 4b; as $\alpha$ gets closer to $\pi$, the peak will move to infinity: this corresponds to a perfect lens as defined in Ref. [2].

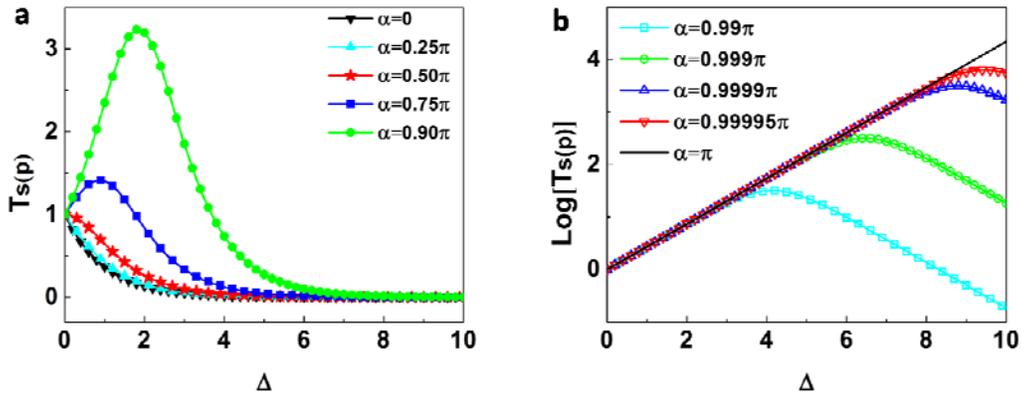

Fig. 4 (a) The amplification of evanescent waves for different $\alpha$. (b) Same as (a) but on a log scale as $\alpha$ is very close to $\pi$.

In conclusion, we find that some of the conjugate metamaterials we describe may serve as a subwavelength-resolution lens, with a perfect lens as the limiting case. Tiny deviations from $\varepsilon = \mu = -1$ will produce imperfections in the imaging functionality. Nevertheless, CMs add to the catalog of metamaterials and deserve further study.

**Acknowledgements**
This work is supported by the National Science Foundation of China for Excellent Young Scientists

(grant no. 61322504), the Foundation for the Author of National Excellent Doctoral Dissertation of China (grant no. 201217), and the Priority Academic Program Development (PAPD) of Jiangsu Higher Education Institutions.